\documentclass[aps,prb,superscriptaddress, reprint]{revtex4-1}
\usepackage{graphicx}
\usepackage{dcolumn}
\usepackage{bm}
\usepackage{amsmath}

\usepackage{color}
\usepackage{bm,graphicx,hyperref}
\hypersetup{%
  breaklinks = {true},
  citecolor = {blue},
  colorlinks = {true},
  linkcolor = {red},
}

\begin{document}

\title{Graphene-mediated interaction between adsorbed impurities}

\author{Keian Noori}
\affiliation{Centre for Advanced 2D Materials, 
			National University of Singapore, 
			6 Science Drive 2, 117546, Singapore}
	
\author{Hillol Biswas}
\affiliation{Department of Physics, National University of
	Singapore, 2 Science Drive 3, 117542, Singapore}
	
\author{Su Ying Quek}
\affiliation{Centre for Advanced 2D Materials, 
			National University of Singapore, 
			6 Science Drive 2, 117546, Singapore}
\affiliation{Department of Physics, National University of
	Singapore, 2 Science Drive 3, 117542, Singapore}

\author{Aleksandr Rodin}
\affiliation{Yale-NUS College, 16 College Avenue West, 138527, Singapore}
\affiliation{Centre for Advanced 2D Materials, 
			National University of Singapore, 
			6 Science Drive 2, 117546, Singapore}
	
\date{\today}

\begin{abstract}

Interaction between adsorbed atoms in graphene is studied using a combination of DFT and the path integral formalism. Our results reveal a complex non-monotonic interaction profile. We show that the strength and sign of the interaction are dictated by the arrangement of impurities, as well as the system doping. These findings can be used to interpret the complex behavior of impurities in experimentally realized systems, as well as other classes of impurities, such as C substitutions in graphene.

\end{abstract}

\maketitle

\section{Introduction}
\label{sec:Introduction}

The problem of graphene-mediated interaction between impurities has been the topic of a number of studies in recent years. In their pioneering work, the authors of Ref.~\onlinecite{Shytov2009} predicted that adatoms hosted on the same sublattice repel while those on the opposite sublattice attract. More recently~\cite{LeBohec2014, Agarwal2019}, however, it has been shown that while the sign of the interaction does depend on the sublattice arrangement, there are also other factors which determine the nature of the interaction. It has been demonstrated, for example, that the interaction can change sign with the separation between the impurities and their on-site energies. In addition, \emph{ab initio} calculations have predicted attraction between adsorbed atoms regardless of the host sublattice.~\cite{Gonzalez-Herrero2016asc}

The purpose of the present work is to explore this non-trivial interaction in a realistic system. The impurities considered here are hydrogen adsorbates, similar to the experimental setup of Ref.~\onlinecite{Gonzalez-Herrero2016asc}. We use two complementary approaches in this study. At small impurity separations, we employ density functional theory (DFT) as it is provides accurate ground-state charge densities and total energy differences for our model system. With increasing distance between the impurities, DFT becomes prohibitively expensive due to large unit cells. Therefore, we turn to the path integral formalism to describe the system. By using the entire Brillouin zone in our path integral calculations we ensure that the symmetries of the system are respected, allowing for a direct comparison between the two approaches to demonstrate both qualitative and quantitative agreement. We find that the presence of a hydrogen impurity induces a non-trivial oscillatory charge on graphene. Further, for systems of two impurities, we demonstrate that the sign of the interaction depends not only on the impurity arrangement, but also on the system doping. The key novelty of the path integral approach described here is its ability to treat a general arrangement of an arbitrary number of impurities without resorting to the Dirac cone approximation. This allows for the use of numerical energy minimization algorithms to study dispersion or aggregation of multiple impurities.

The paper is organized as follows. In Sec.~\ref{sec:DFT_Results}, we provide the results obtained from DFT. Section~\ref{sec:Analytical_Results} uses the path integral formalism to construct a model used to compute the charge density variation as well as the impurity interaction energy. Finally, the summary can be found in Sec.~\ref{sec:Summary}.

\section{DFT Results}
\label{sec:DFT_Results}
\begin{figure}[hbtp]
	\centering
	\includegraphics[width=\columnwidth]{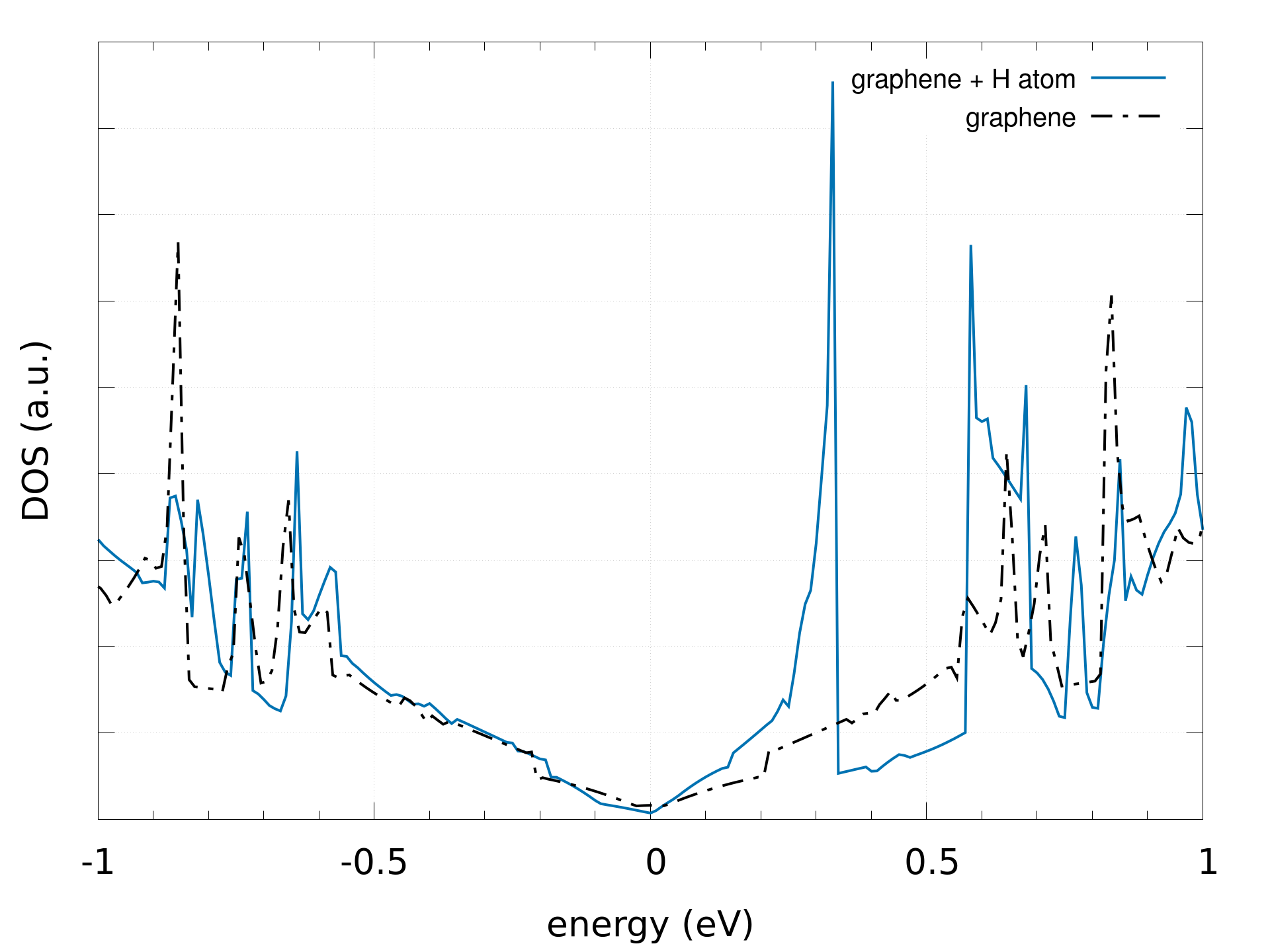}
	\caption{DFT DOS for both pristine (dashed black line) and single H-impurity (solid blue line) graphene. The addition of the H adatom breaks the \emph{sp$^2$} bonding of the underlying C atom and creates an impurity state $\sim$~0.3 eV above the Dirac point of pristine graphene.}
	\label{fig:DOS:DFT}
\end{figure}

We begin by exploring the effect of the adsorption of a single hydrogen (H) atom impurity on a $12 \times 12$ supercell of pristine, planar (unrelaxed) graphene. When a H atom is added to the system, it binds to an individual carbon (C) atom by forming $\sigma$ bonds with the C $p_z$ and $s$ orbitals. The coupling between carbon's $p_z$ orbital and hydrogen's $s$ orbital leads to level repulsion. This produces a state with an energy above the Dirac point, as shown in Fig.~\ref{fig:DOS:DFT}. The sharpness of the peak is the consequence of the weak coupling between the host C and its neighbors. In effect, the host atom acts as a vacancy in graphene without the lattice reconstruction due to a missing atom.~\cite{Soriano2011}

\begin{figure}[hbtp]
	\centering
	\includegraphics[width=\columnwidth]{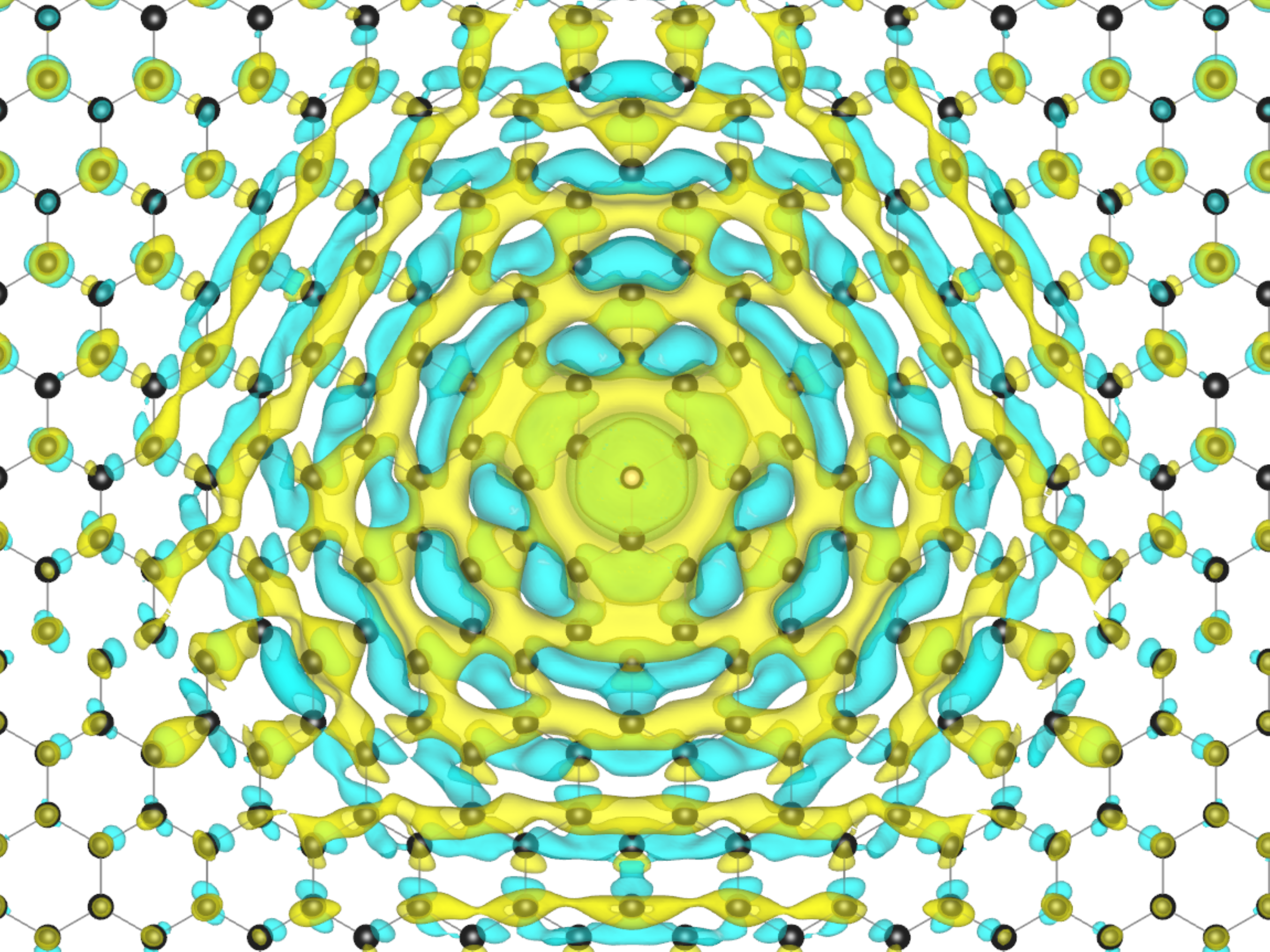}
	\caption{Induced charge density in undoped graphene (black atoms) upon adsorption of a single H adatom (white atom). Yellow regions represent a net accumulation of electrons, while blue regions represent a net deficit of electrons. An illustrative isosurface value of $3\times10^{-5}$ electron/bohr$^{3}$ is used.}
	\label{fig:Charge:DFT}
\end{figure}

The adsorbed H adatom modifies the charge density of pristine graphene such that the induced density oscillates in a concentric manner around the impurity, as shown in Fig.~\ref{fig:Charge:DFT}. Notably, the period of this charge oscillation is much shorter than is expected of the well-known Friedel oscillations. We observe a sublattice dependence, which is more apparent further away from the impurity. Charge depletion/accumulation tends to occur at sites on the opposite/same sublattice as the impurity.

\begin{figure*}[hbtp]
	\includegraphics[width = 3.5in]{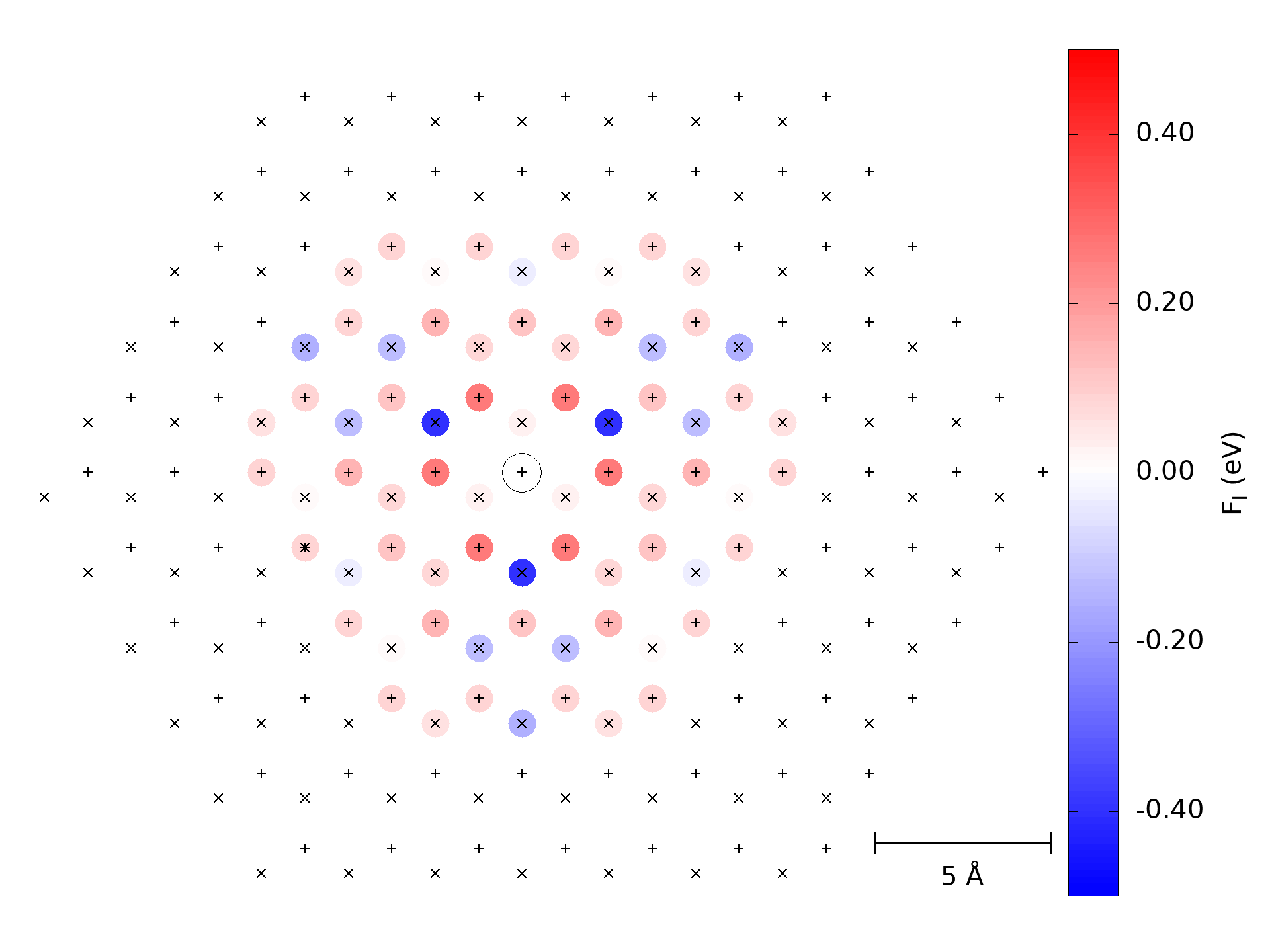}
	\includegraphics[width = 3.5in]{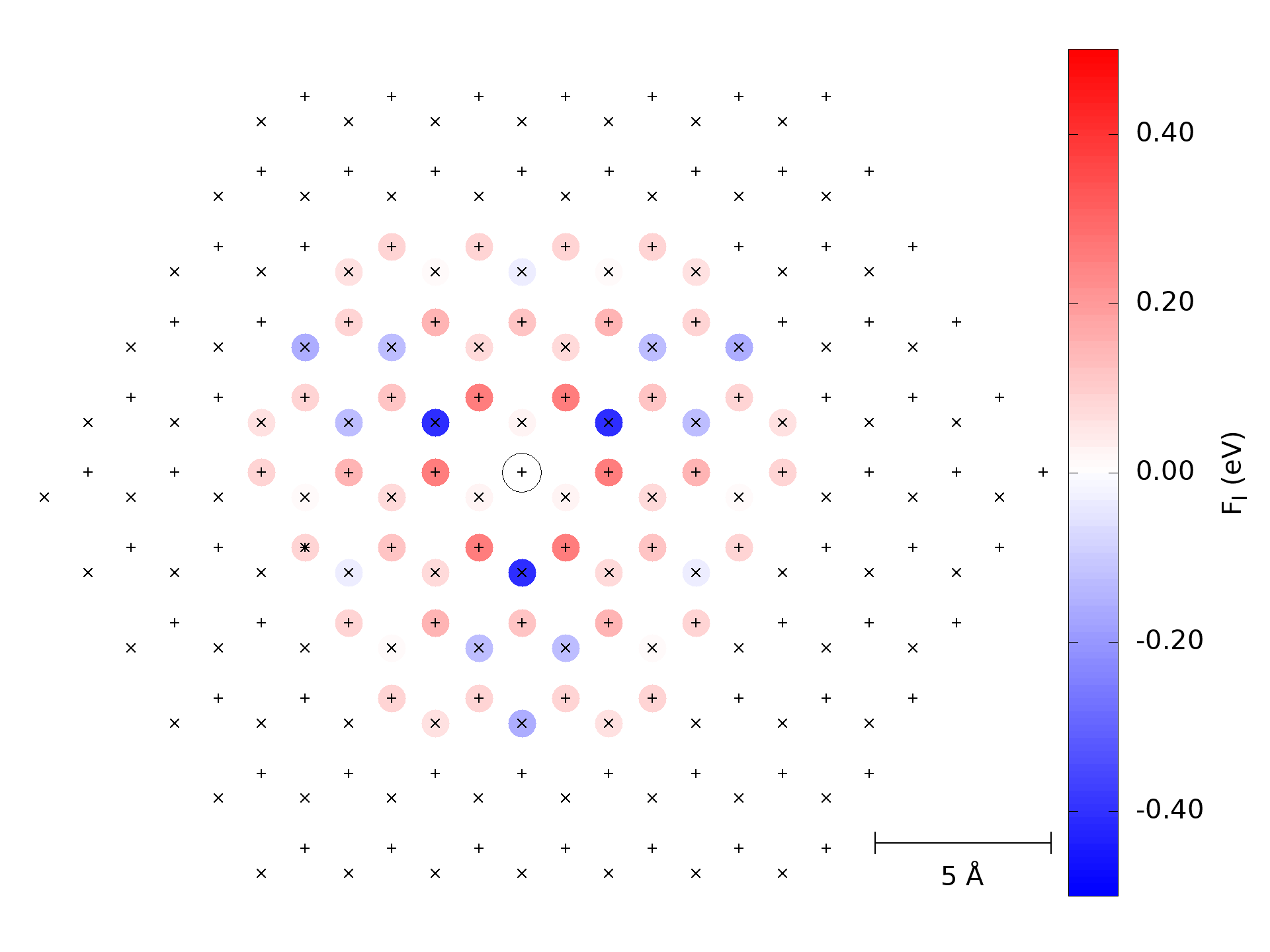}
	\includegraphics[width = 3.5in]{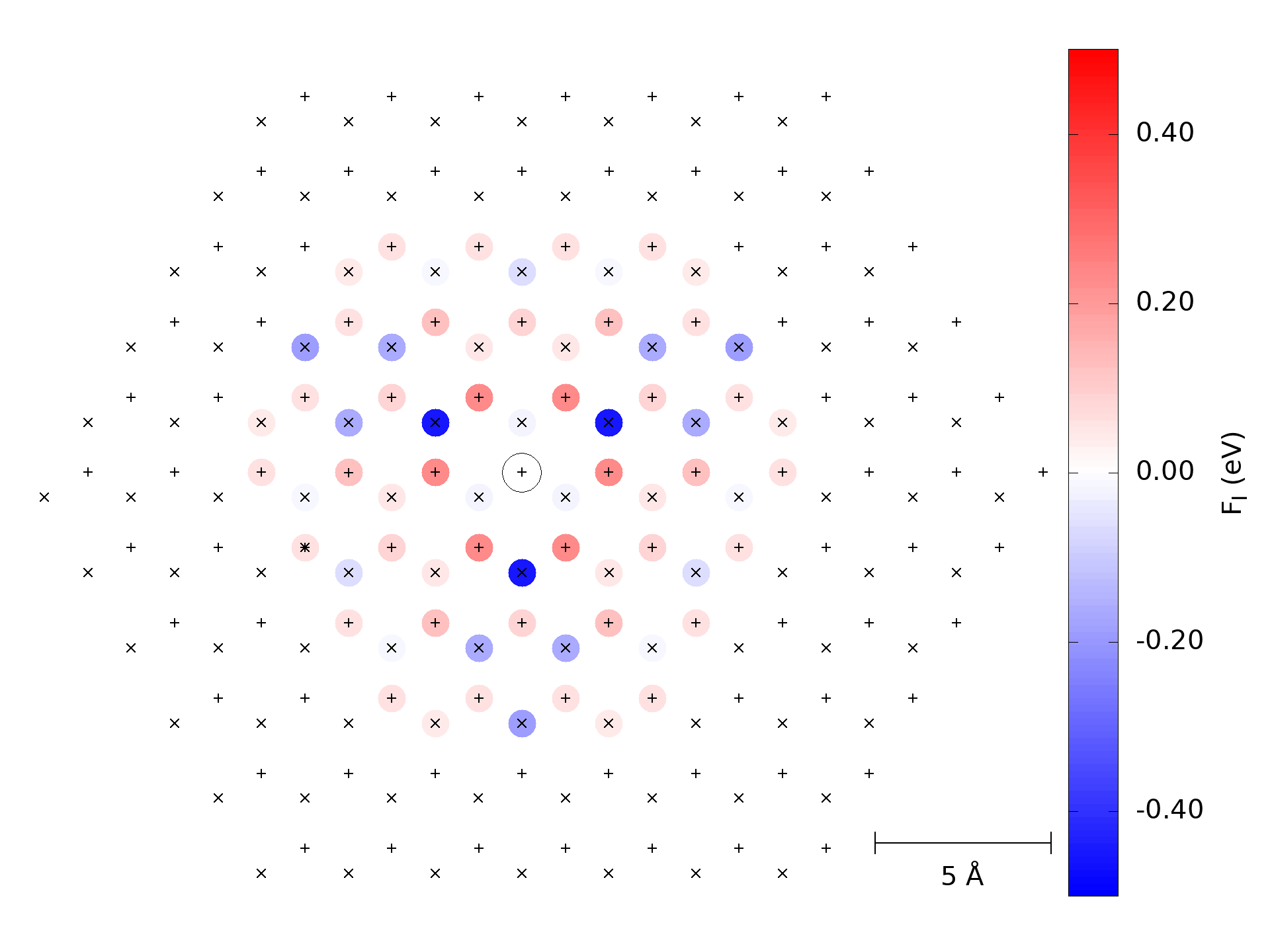}
	\includegraphics[width = 3.5in]{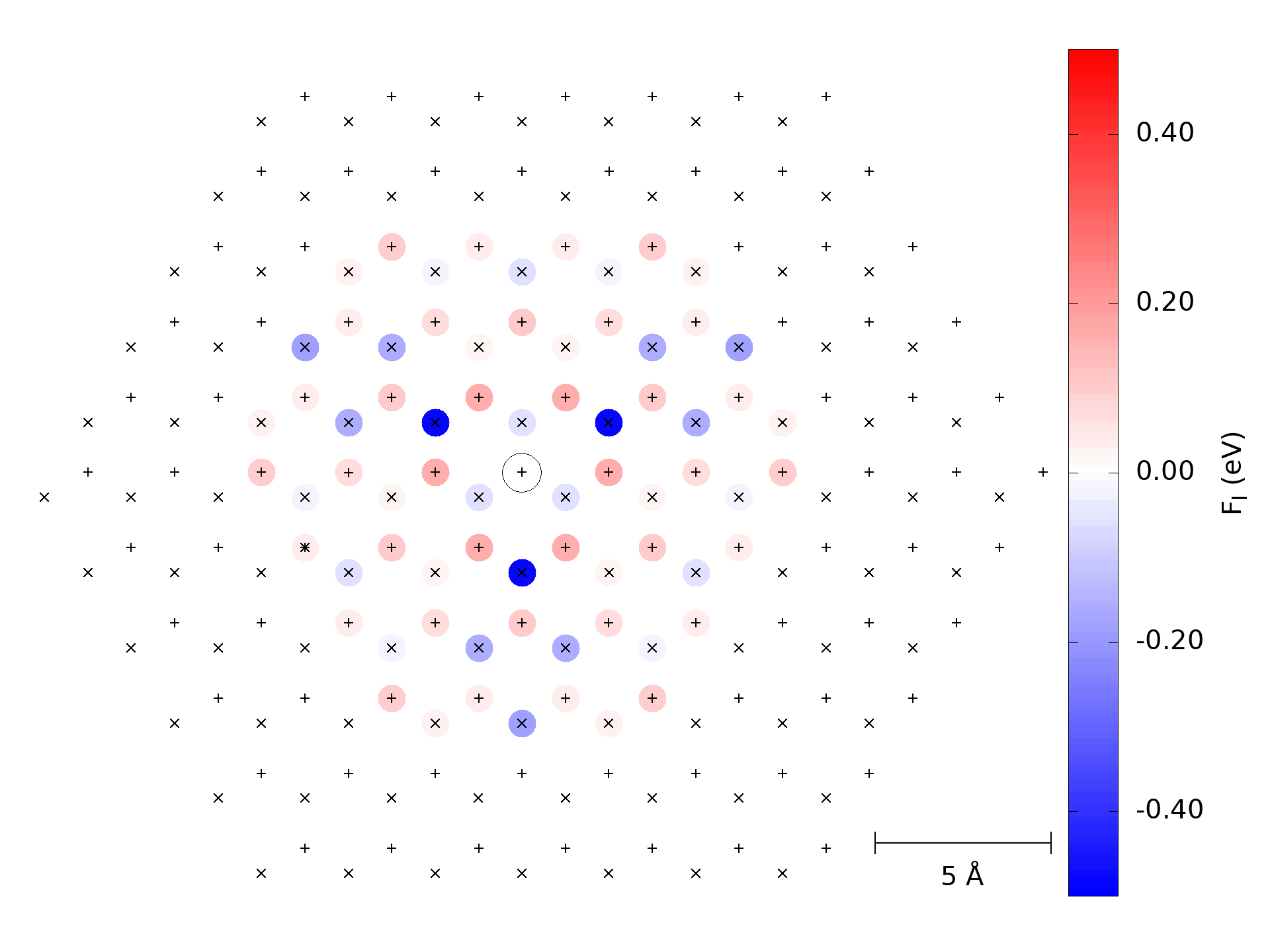}
	\caption{Interaction energy, $F_{I}$, between two H impurities adsorbed above graphene in a $12 \times 12$ supercell. One impurity, $H^{(0)}$, is fixed at the center of the graphene supercell (marked with a circle), while the second, $H^{(l)}$, is variously placed above a second C atom. The systems in all four panels are identical aside from their doping levels. \emph{Top left:} undoped graphene. \emph{Top right:} graphene doped with 0.01 electron/unit cell, corresponding to $\sim 1.3 \times 10^{11}$ electron/cm$^{2}$. \emph{Bottom left:} graphene doped with 0.10 electrons/unit cell, corresponding to $\sim 1.3 \times 10^{12}$ electron/cm$^{2}$. \emph{Bottom right:} graphene doped with 1.00 electrons/unit cell, corresponding to $\sim 1.3 \times 10^{13}$ electron/cm$^{2}$. The A and B graphene sublattices are indicated by $\mathbf{+}$ and $\mathbf{x}$ markers, respectively.}
	\label{fig:Interaction:DFT}
\end{figure*}

In order to compute the interaction energy between two impurities on graphene, we place two H atoms above different C atoms on the graphene supercell. We fix one H impurity above the central C atom, while variously moving the second H atom to one of 72 different positions around this central point. This scheme is shown pictorially in Fig.~\ref{fig:Interaction:DFT}. The interaction energy is computed as
\begin{equation}
	F_I^{(l)} = E_{b}^{H^{(0)}H^{(l)}} - E_{b}^{H^{(0)}} - E_{b}^{H^{(l)}},
	\label{eqn:F_I_dft}
\end{equation}
where $E_{b}^{H^{(0)}H^{(l)}}$ is the binding energy of the two-impurity system with H adsorbates on C atoms at positions $0$ and $l$, while $E_{b}^{H^{(0)}}$ and $E_{b}^{H^{(l)}}$ represent the binding energies of one-impurity systems with the H adsorbate at position $0$ and $l$, respectively.
The interaction energy, shown in Fig.~\ref{fig:Interaction:DFT}, decays non-monotonically as the distance between the two H impurities increases. When both H adatoms are close to each other, the magnitude of $F_{I}$ peaks at $\sim$~0.4 eV, while at further separations the magnitude of $F_{I}$ tends toward a converged value of $\sim$~0.05 eV. The scale of this interaction is significantly larger than that predicticted in an earlier theoretical work~\citep{Shytov2009} that suggests an interaction energy less than 10 meV for the same concentration of adatoms.

As with the induced charge density redistribution by a single H impurity, the two-impurity interaction oscillates on a length scale much shorter than the Friedel oscillations. Further, the interaction also depends on the sublattice, with the interaction energy rising when both adatoms are on the same sublattice, and decreasing when they are on different sublattices. We note that we expect a small quantitative error in our values of $F_{I}$ due the effect of the periodic images. We stress, however, that these periodic effects do not alter the qualitative picture, nor do they change the order of magnitude of the interactions. In effect, the periodicity adds a constant offset value to $F_{I}$ since the adatoms in each cell feel the impact of those adatoms in the repeated cells, pushing $F_{I}$ away from zero even for large impurtiy separations. This effect will decrease as the size of the supercell is increased. We also explore the effects of spin polarization and structural relaxation and find both to be qualtitatively and quantitatively insignificant. Further details regarding these effects are found in Appendix~\ref{sec:CompDetails}.

We examine the effect of shifting the chemical potential, $\mu$, by extracting 0.01, 0.10 and 1.00 electrons per cell (see Fig.~\ref{fig:Interaction:DFT}), corresponding to a p-type doping of approximately $1.3 \times10^{11}$, $1.3 \times10^{12}$, and $1.3 \times10^{13}$ electron/cm$^2$, respectively. There is a small but clear quantitative effect on $F_{I}$ as $\mu$ is shifted, indicating that the position of the Fermi level affects the total energy by way of the hybridized graphene--adatom states. At the length scales explored in our DFT model, however, we do not observe a qualitative shift in the behavior with respect to the undoped case.

In order to analyze the effects of impurities at larger distances, as well as to gain a deeper insight into the mechanisms leading to the oscillations and $\mu$-dependence seen above, we turn to a path integral formalism.

\section{Analytical Results}
\label{sec:Analytical_Results}

\subsection{Model}
\label{sec:Model}

The starting point for the analytical model is the familiar graphene Hamiltonian with nearest-neighbor hopping
\begin{equation}
	\hat{H}_G =
	\sum_{\mathbf{q}\in \mathrm{BZ}}
	\begin{pmatrix}
		a^\dagger_{\mathbf{q}},b^\dagger_{\mathbf{q}}
	\end{pmatrix}
	\begin{pmatrix}
		-\mu             & -tf_\mathbf{q}
		\\
		-tf^*_\mathbf{q} & -\mu
	\end{pmatrix}
	\begin{pmatrix}
		a_{\mathbf{q}} \\ b_{\mathbf{q}}
	\end{pmatrix}
	\,,
	\label{eqn:H_G}
\end{equation}
where $t$ is the hopping integral, $f_\mathbf{q} = 1 + e^{i\mathbf{q}\cdot\mathbf{d}_1} + e^{i\mathbf{q}\cdot\mathbf{d}_2}$, and $\mathbf{d}_{1/2} = d \left(\pm 1,\sqrt{3}\right)/2$ are the lattice vectors. To keep the notation as concise as possible, we will use $c^\dagger_\mathbf{q} = \begin{pmatrix}
		a^\dagger_{\mathbf{q}},b^\dagger_{\mathbf{q}}
	\end{pmatrix}$.

To describe the hydrogen impurity, we make a few approximations. First and foremost, we neglect the distant $\sigma$ bands of graphene. In addition, we ignore the internal dynamics of the C-H dimer. Instead, we will treat it as a carbon atom with a modified $p_z$ energy and a reduced nearest-neighbor hopping parameter. This quasi-isolated orbital produces a peak in the DOS like the one seen in Fig.~\ref{fig:DOS:DFT}.

For reasons which will become apparent shortly, it is convenient to write the impurity-driven part of the Hamiltonian as
\begin{equation}
	\hat{H}_I = \sum_{jk} c^\dagger_{\mathbf{R}_j}
	I_j \Delta_{jk}I^T_k c_{\mathbf{R}_k}\,.
	\label{eqn:H_I}
\end{equation}
To elucidate the notation used here, we start with $j = k$ to get terms like $c^\dagger_{\mathbf{R}_j} I_j \Delta_{jj}I^T_j c_{\mathbf{R}_j}$. These terms describe the change of the on-site energy at the $j$th atom given by $\Delta_{jj}$. The location of the unit cell hosting this atom is $\mathbf{R}_j$ and
$I_j^T =
	\begin{pmatrix}
		1 & 0
	\end{pmatrix}$
or
$\begin{pmatrix}
		0 & 1
	\end{pmatrix}$ (which we denote as $A$ and $B$, respectively) depending on the atom's sublattice. For $j\neq k$, Eq.~\eqref{eqn:H_I} describes the modified hopping integrals between the $j$th and $k$th atoms. For a system composed of $N$ unit cells, the Fourier-space version of Eq.~\eqref{eqn:H_I} is
\begin{equation}
	\hat{H}_I = \frac{1}{N}\sum_{\mathbf{qq}'}\sum_{jk} c^\dagger_{\mathbf{q}}	I_{j\mathbf{q}} \Delta_{jk}I^\dagger_{k\mathbf{q}'}
	c_{\mathbf{q}'}
	\label{eqn:H_I_Fourier}
\end{equation}
with $I_{j\mathbf{q}}= I_j e^{-i\mathbf{q}\cdot\mathbf{R}_j}$.

Equations~\eqref{eqn:H_G} and \eqref{eqn:H_I_Fourier} can be combined into the imaginary time action:
\begin{align}
	S
	                            & =
	\sum_{\mathbf{q}\omega_n}
	\bar\Phi_{\mathbf{q}\omega_n}
	\left(-G^{-1}_{\mathbf{q}\omega_n}\right)
	\Phi_{\mathbf{q}\omega_n}
	\nonumber
	\\
	                            &
	+
	\frac{1}{N}
	\sum_{\mathbf{qq}'\omega_n}
	\bar\Phi_{\mathbf{q}\omega_n}
	I_{\mathbf{q}}
	\Delta
	I^\dagger_{\mathbf{q}'}
	\Phi_{\mathbf{q}'\omega_n}\,,
	\label{eqn:S}
	\\
	G^{-1}_{\mathbf{q}\omega_n} & =G^{-1}_\mathbf{q}\left(i\omega_n + \mu\right)=
	\begin{pmatrix}
		i\omega_n+\mu   & tf_\mathbf{q}
		\\
		tf^*_\mathbf{q} & i\omega_n+\mu
	\end{pmatrix}
	\label{eqn:G}\,,
\end{align}
where $G^{-1}_{\mathbf{q}\omega_n} $ is the inverse of the pristine graphene Green's function and $I_\mathbf{q} = \begin{pmatrix}
		I_{1\mathbf{q}} & I_{2\mathbf{q}} & \dots
	\end{pmatrix}$. Exponentiating $-S$ and integrating over all fields yields the partition function
\begin{align}
	\mathcal{Z}
	                                        & =
	\prod_{\omega_n}\left|-\beta\hat{\mathcal{G}}^{-1}_{\omega_n}\right|\,,
	\label{eqn:Z}
	\\
	\mathcal{G}^{-1}_{\mathbf{qq}'\omega_n} & =  G^{-1}_{\mathbf{q}\omega_n}\delta_{\mathbf{qq}'}
	-
	\frac{1}{N}
	I_{\mathbf{q}}
	\Delta
	I^\dagger_{\mathbf{q}'}	\,,
	\label{eqn:G_full_Inv}
\end{align}
where $\hat{\mathcal{G}}^{-1}_{\omega_n}$ is the inverse of the full Green's function. Defining $\mathcal{I}$ as a column vector of $I_\mathbf{q}$ gives $\mathcal{G}^{-1}_{\omega_n} = G^{-1}_{\omega_n}-
	N^{-1}
	\mathcal{I}
	\Delta
	\mathcal{I}^\dagger$ so that
\begin{equation}
	\hat{\mathcal{G}}_{\omega_n}
	=
	\hat{G}_{\omega_n}
	+
	\hat{G}_{\omega_n} \mathcal{I}\frac{\Delta}{N}
	\left[1-\frac{\mathcal{I}^\dagger\hat{G}_{\omega_n} \mathcal{I}}{N}\Delta\right]^{-1}
	\mathcal{I}^\dagger\hat{G}_{\omega_n}\,.
	\label{eqn:G_full}
\end{equation}

In addition, the free energy, given by $F = -T\ln\mathcal{Z}$, becomes
\begin{align}
	F
	= & -  T\sum_{\omega_n}\ln\left|-\beta\hat{G}^{-1}_{\omega_n}
	\right|
	\nonumber
	\\
	  & -  T\sum_{\omega_n}
	\ln\left|
	1
	-\frac{\mathcal{I}^\dagger\hat{G}_{\omega_n}\mathcal{I}}{N}\Delta
	\right|\,.
	\label{eqn:F}
\end{align}
Here, the first term is the free energy of pristine graphene, while the second term gives the free energy contribution due to the impurities.

The quantity that appears both in the free energy and the Green's function is
\begin{align}
	  & \frac{\mathcal{I}^\dagger\hat{G}_{\omega_n} \mathcal{I}}{N} =\frac{1}{N}\sum_\mathbf{q}I^\dagger_\mathbf{q} \hat{G}_{\mathbf{q}\omega_n} I_\mathbf{q}
	\nonumber
	\\
	= &
	\frac{1}{N}\sum_\mathbf{q}
	\begin{pmatrix}
		I^\dagger_{1\mathbf{q}} \hat{G}_{\mathbf{q}\omega_n}I_{1\mathbf{q}}
		       &
		I^\dagger_{1\mathbf{q}} \hat{G}_{\mathbf{q}\omega_n} I_{2\mathbf{q}}
		       &
		\dots
		\\
		I^\dagger_{2\mathbf{q}} \hat{G}_{\mathbf{q}\omega_n} I_{1\mathbf{q}}
		       &
		I^\dagger_{2\mathbf{q}} \hat{G}_{\mathbf{q}\omega_n} I_{2\mathbf{q}}
		       &
		\dots
		\\
		\vdots & \vdots & \ddots
	\end{pmatrix}\,.
	\label{eqn:Scattering}
\end{align}
The summation over the momentum can be written as
\begin{equation}
	\frac{1}{N}\sum_\mathbf{q} I^\dagger_{j\mathbf{q}} \hat{G}_{\mathbf{q}\omega_n} I_{k\mathbf{q}}
	=
	I^T_{j} \Xi_{\omega_n}^{\mathbf{R}_j - \mathbf{R}_k} I_{k}\,,
\end{equation}
where $\Xi^\mathbf{R}_{\omega_n} = \Xi^\mathbf{R}\left(i\omega_n + \mu\right)$ and
\begin{equation}
	\Xi^\mathbf{R}\left(z\right) = \frac{1}{N}\sum_\mathbf{q} G_\mathbf{q}\left(z\right)e^{i\mathbf{q}\cdot\mathbf{R}}\,.
	\label{eqn:Xi}
\end{equation}
To make the expression more compact, Eq.~\eqref{eqn:Scattering} can be written as $\mathbf{I}^T \mathbf{\Xi}\mathbf{I}$, where $\mathbf{I}$ is a diagonal matrix of $I$ and $\mathbf{\Xi}$ is the matrix of $\Xi_{\omega_n}$. For the calculation of $\Xi_{\omega_n}^{\mathbf{R}_j - \mathbf{R}_k}$, refer to Appendix~\ref{sec:Graphene_Propagator}.

\subsection{Local Electronic Density}
\label{sec:Local_Electronic_Density}

The local electronic density is calculated using the Green's function from Eq.~\eqref{eqn:G_full}. To do so, recall that the electron expectation number at the $j$th component of the $\Phi$ vector in the unit cell at $\mathbf{R}$ is
\begin{align}
	\rho^j_{\mathbf{R}} & = \frac{1}{\beta}\sum_{\omega_n} \langle \bar{\phi}^j_{\mathbf{R}\omega_n}\phi^j_{\mathbf{R}\omega_n}\rangle
	\nonumber
	\\
	                    & =\frac{1}{\beta}\sum_{\mathbf{qq}'\omega_n} \langle \bar{\phi}^j_{\mathbf{q}\omega_n}\phi^j_{\mathbf{q}'\omega_n}\rangle \frac{e^{i\left(\mathbf{q}'-\mathbf{q}\right)\cdot\mathbf{R}}}{N}\,.
	\label{eqn:rho}
\end{align}
The required correlation function $ \langle \bar{\phi}^j_{\mathbf{q}\omega_n}\phi^j_{\mathbf{q}'\omega_n}\rangle$ is given by the diagonal elements of the $\mathcal{G}_{\mathbf{q}'\mathbf{q}\omega_n} $ blocks of the full bulk Green's function. These blocks include two parts: the pristine bulk contribution and the term containing the adatom effects. Since we are interested in the adsorbate-induced charge variation, we focus on the second term and introduce
\begin{align}
	\delta \mathcal{G}_{\mathbf{R}\omega_n} = &
	\sum_{\mathbf{qq}'}
	\hat{G}_{\mathbf{q}'\omega_n}I_{\mathbf{q}'}\Delta
	\left[1-\mathbf{I}^T\mathbf{\Xi I}\Delta\right]^{-1}
	I^\dagger_{\mathbf{q}}\hat{G}_{\mathbf{q}\omega_n}
	\label{eqn:Delta_G}
	\\
	\times                                    &
	\frac{e^{i\left(\mathbf{q}'-\mathbf{q}\right)\cdot\mathbf{R}}}{N^2}
	\nonumber
	\\
	=                                         &
	\begin{pmatrix}
		\Xi_{\omega_n}^{\mathbf{R} - \mathbf{R}_1}
		 &
		\dots
	\end{pmatrix}\mathbf{I}\Delta
	\left[1-\mathbf{I}^T\mathbf{\Xi I}\Delta\right]^{-1}
	\mathbf{I}^T
	\begin{pmatrix}
		\Xi_{\omega_n}^{\mathbf{R}_1 - \mathbf{R}}
		 & \\
		\vdots
	\end{pmatrix}
	\,,
	\nonumber
\end{align}
which is the impurity-induced correction to the real-space graphene correlation function. The diagonal elements of $\delta \mathcal{G}_{\mathbf{R}\omega_n}$ summed over $\omega_n$ give the correction to the electronic density for the corresponding atom of the unit cell at $\mathbf{R}$:
\begin{align}
	\delta \rho_{\mathbf{R}}^s
	 & =
	\frac{1}{\beta}\sum_{\omega_n}
	\begin{pmatrix}
		s^T\,\Xi_{\omega_n}^{\mathbf{R} - \mathbf{R}_1}I_1
		 &
		\dots
	\end{pmatrix}
	\nonumber
	\\
	& \times
	\Delta
	\left[1-\mathbf{I}^T\mathbf{\Xi I}\Delta\right]^{-1}
	\begin{pmatrix}
		I_1^T\Xi_{\omega_n}^{\mathbf{R}_1 - \mathbf{R}}s
		\\
		\vdots
	\end{pmatrix}
	\,,
	\label{eqn:Delta_rho}
\end{align}
where $s = A$ or $B$.
\begin{figure}
	\includegraphics[width = 0.4\textwidth]{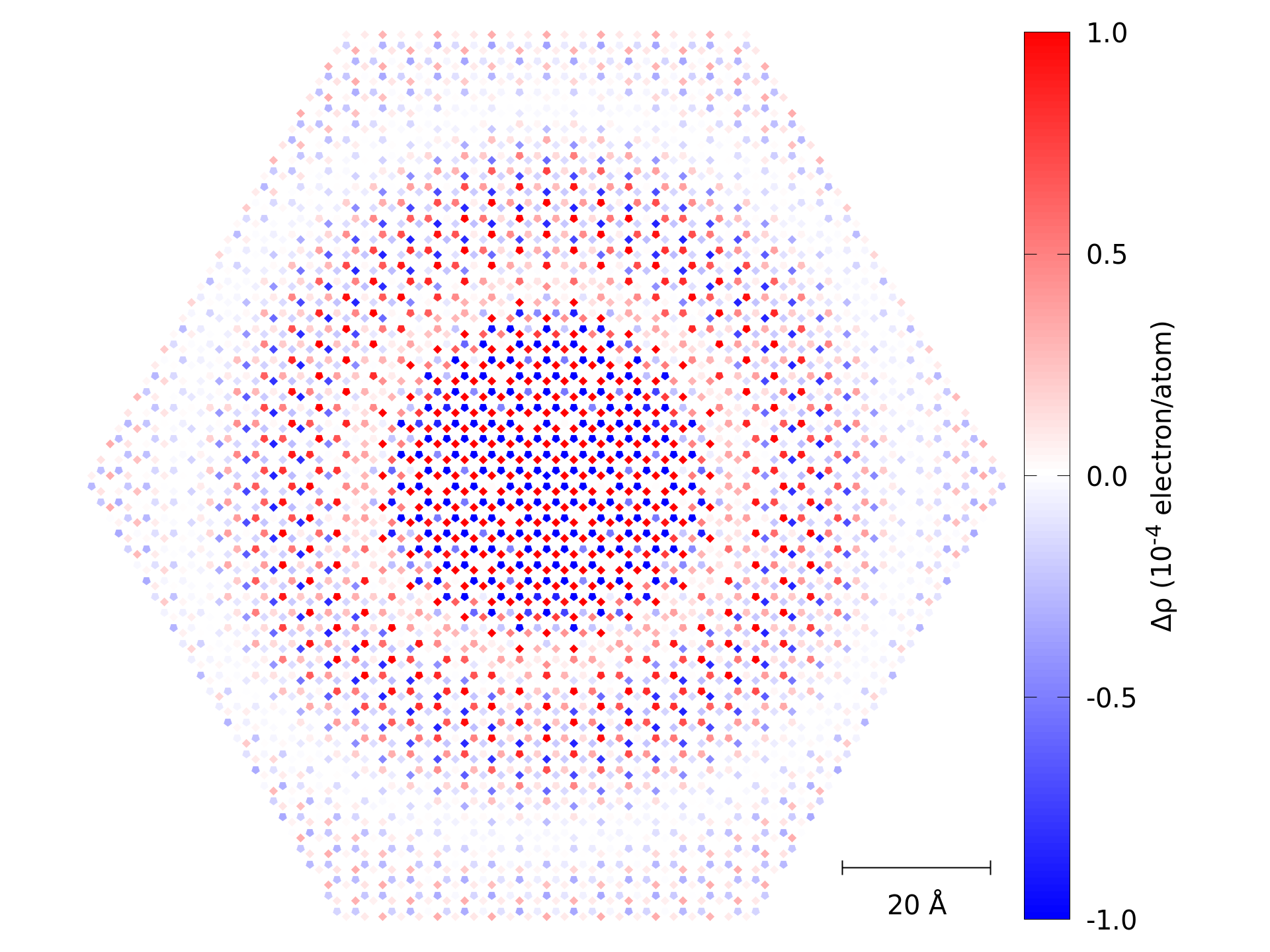}
	\includegraphics[width = 0.4\textwidth]{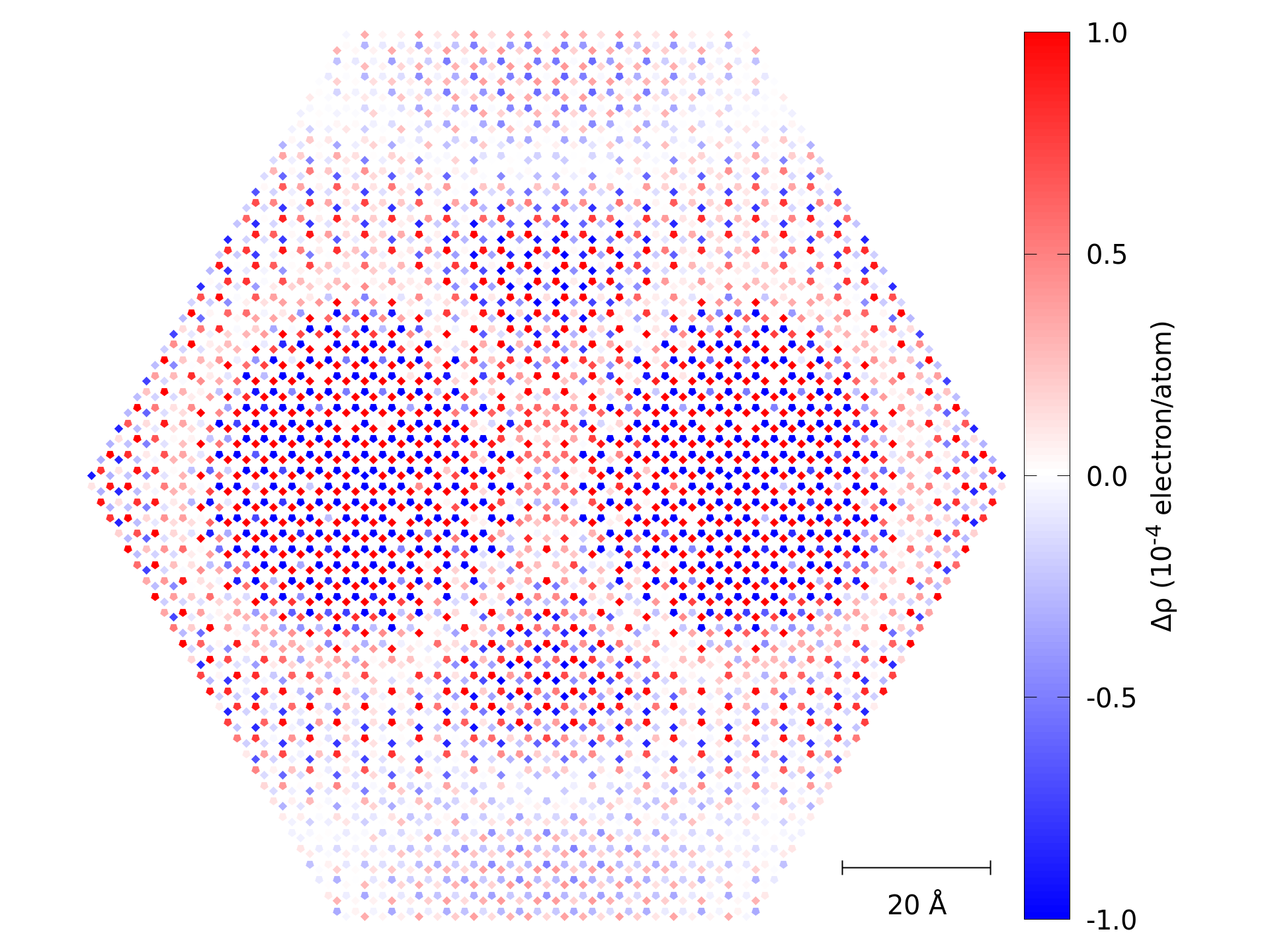}
	\includegraphics[width = 0.4\textwidth]{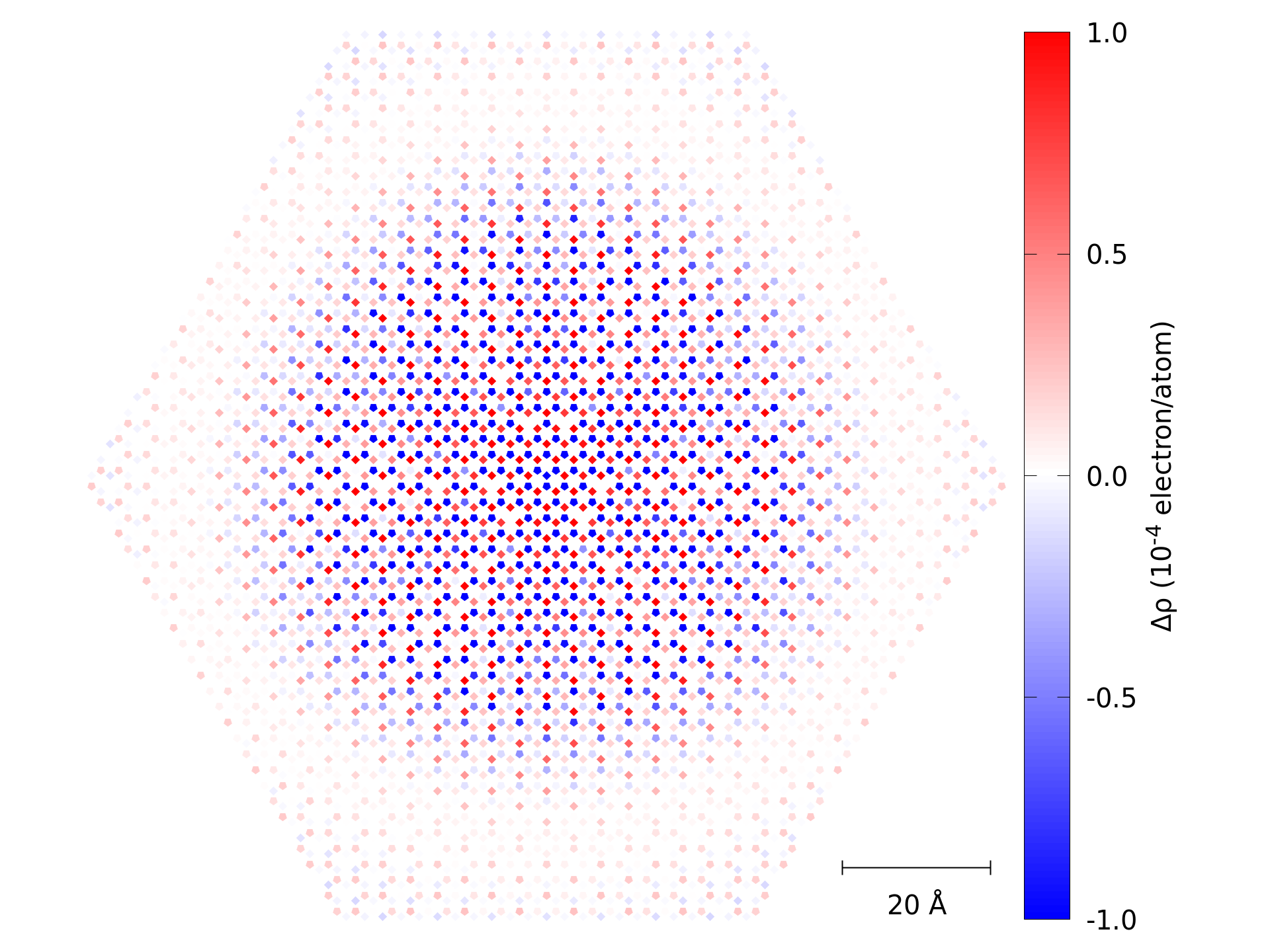}
	\caption{Impurity-induced electron density variation $\Delta \rho$ for $\mu = -0.4$eV (top two panels) and $\mu = -0.2$eV (bottom panel). Two concentric rings where the sublattices have the same sign of $\Delta \rho$ are discernible in the top panel. The rings correspond to the maximum (inner ring) and the minimum (outer ring) of the Friedel oscillations. The difference in their radii ($\approx 25$ \AA) is half of the Friedel period ($52$ \AA\, for $\mu = -0.4$ eV).}
	\label{fig:QFT_rho}
\end{figure}
Using Eq.~\eqref{eqn:Delta_rho}, we compute the electronic density variation for several scenarios and plot it in Fig.~\ref{fig:QFT_rho}. In this calculation, we assume that the host atom becomes completely decoupled from its neighbors~\citep{Soriano2011} since we do not compute the precise value of the modified hopping integral. Allowing finite coupling does not change the qualitative picture and can be done with no additional computational cost if the hopping term is known.

As for any conductor, impurities in graphene are expected to give rise to Friedel oscillations. The peculiarity here is that the oscillatory behavior for the two sublattices is not completely in phase~\cite{Bacsi2010}, as was discussed in Sec.~\ref{sec:DFT_Results}. Because of this, when averaged over several sites, the contribution from the two sublattices partially cancels except for periodic ring-like regions where the oscillations are in-phase, see Fig.~\ref{fig:QFT_rho}. These periodic regions are the Friedel oscillations.

When the chemical potential is set to $\mu = -0.4$ eV (top two panels of Fig.~\ref{fig:QFT_rho}), $k_F \approx 0.06 \mathrm{\AA}^{-1}$. The wavelength of the Friedel oscillations is then $\pi / k_F\approx 52$ \AA. From the top panel of Fig.~\ref{fig:QFT_rho}, one can see that the difference of the radii of two concentric Friedel rings is about 25 \AA. Since this is the separation between the local charge minimum and maximum, it corresponds to one half of the Friedel period, in agreement with the calculation above.

Close to the adatoms, the amplitude of the electronic density oscillations is about $10^{-3}$ electron/atom. At larger distances, it drops to $\sim10^{-4}$ electron/atom. While these numbers may appear small, it is important to put them into the context of graphene doping: $10^{-4}$ electron/atom corresponds to a doping of approximately $4\times 10^{11}$ electron/$\mathrm{cm}^2$.

The Friedel oscillations persist in the two-impurity configuration as well, where the interference modifies their ring-like structure. Reducing the doping to $\mu = -0.2$ eV (bottom panel) increases the oscillation period, as expected.

\subsection{Adatom Interaction Energy}
\label{sec:Adatom_Interaction_Energy}
To obtain the interaction energy between two impurities, we proceed in the same way as in the DFT approach. That is, we calculate the second term of Eq.~\eqref{eqn:F} for both a two-impurity configuration and for a single impurity one. The interaction energy is given by doubling the single-adatom energy and subtracting it from the two-impurity energy.

\begin{figure}
	\includegraphics[width = \columnwidth]{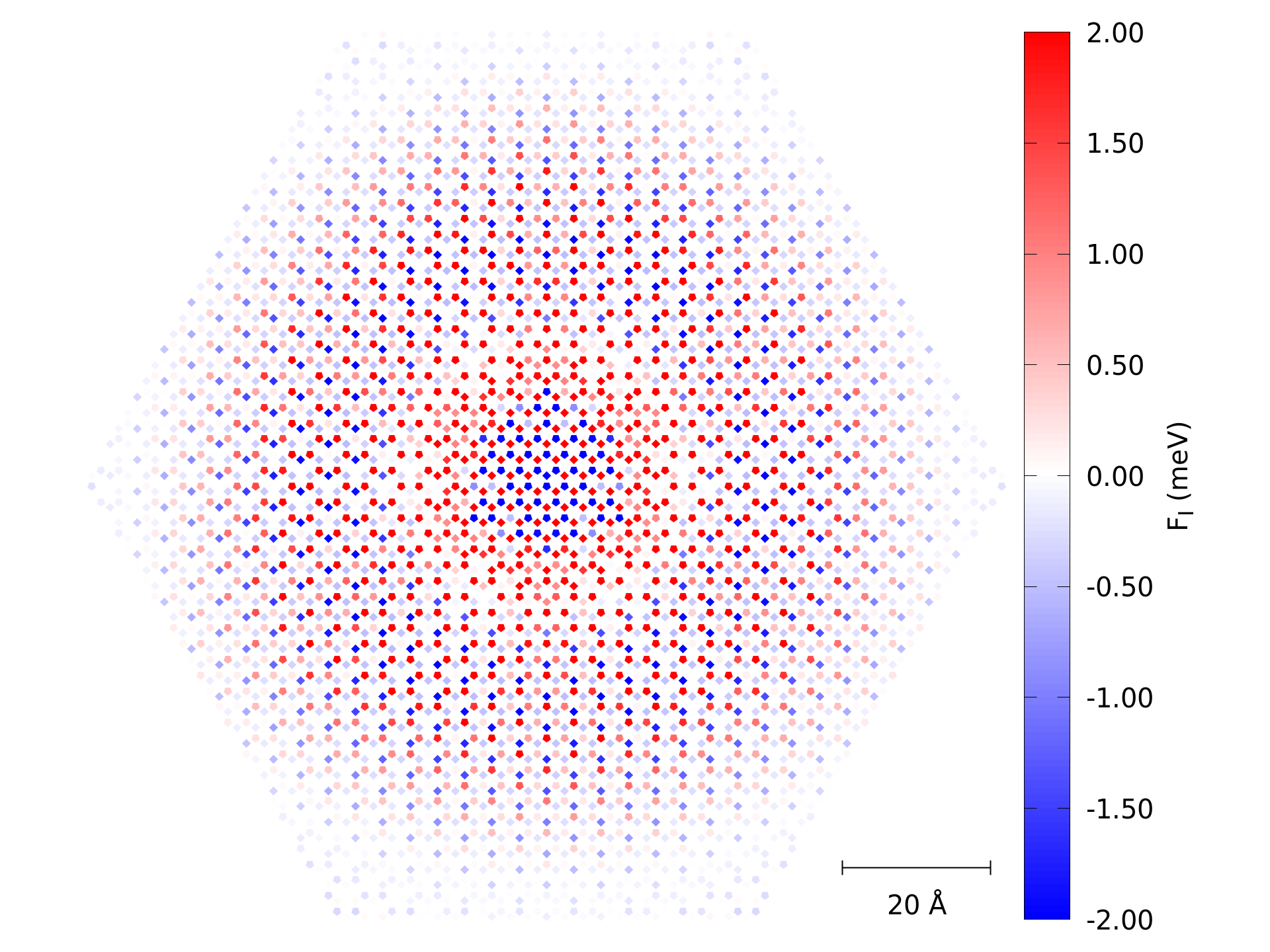}
	\includegraphics[width = \columnwidth]{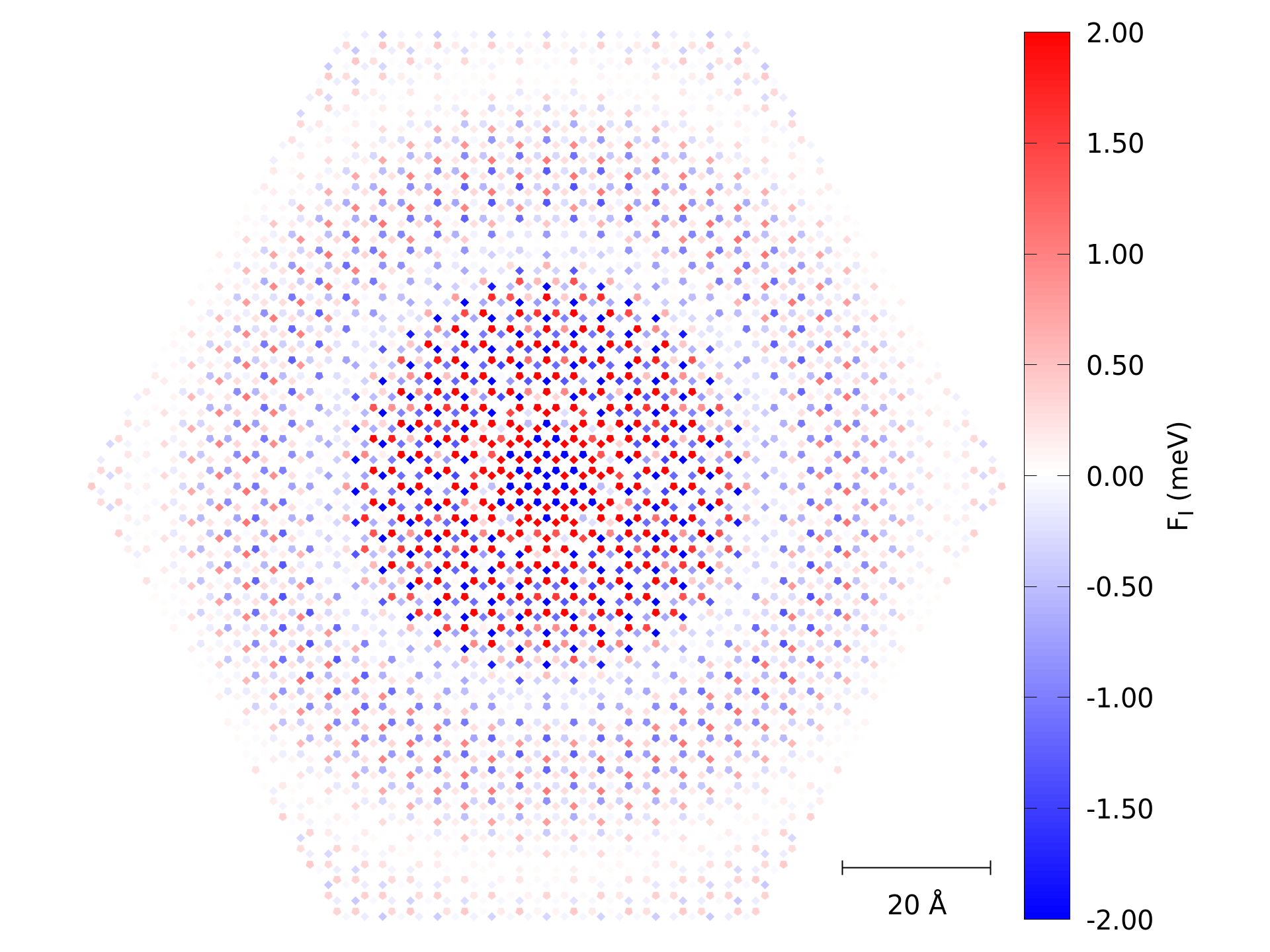}
	\caption{Interaction energy $F_I$ for $\mu = -0.2$ eV (top) and $\mu= -0.4$ eV (bottom). The color scale is saturated to illustrate the oscillatory behavior at large separations.}
	\label{fig:QFT_FI}
\end{figure}

We calculate the interaction energy for the same doping levels that we used to illustrate the Friedel oscillations above (see Fig.~\ref{fig:QFT_FI}). In fact, we observe that the interaction energy exhibits oscillations of the same period as the charge density, albeit shifted by a phase. In agreement with the DFT results, we see that for small separations, the system energy changes sign depending on the location of the impurities, becoming negative/positive when the adsorbates are hosted on the opposite/same sublattice. The rising/lowering of the energy is consistent with the DFT results of Fig.~\ref{fig:Interaction:DFT}, and, following the discussion in Sec.~\ref{sec:DFT_Results}, the sign change is consistent with the DFT energies of a very large system in which the periodic effects are negligible (i.e., the large-separation, converged value of $F_{I}$ goes to zero).

The level of doping is similar to what is commonly observed experimentally and to the values we chose for the DFT calculations. This allows us to investigate the quantitative agreement between the two methods. At very small separations, $F_I$ is of the order of $100$ meV, as seen in the DFT results. In addition, the interaction is stronger when the chemical potential is closer to the impurity level.

\section{Summary}
\label{sec:Summary}
We have performed a detailed study on graphene-mediated impurity interactions using a combined \emph{ab initio} and path integral formalism approach. Our results demonstrate the non-trivial oscillatory behavior of the adatom-induced charge density and impurity interaction. Consitent with previous theoretical analysis, we show that there exists a sublattice dependence to the sign of the interaction. We also demonstrate that the interaction decays non-monotonically, with oscillation in the interaction energy of impurities depending on the their sublattices, distance, and the chemical potential of the system. Our results extend the current understanding of graphene-mediated impurity interactions, and can be used to help understand the complex behavior of impurities in experimentally realized systems~\citep{Gonzalez-Herrero2016asc}. The path integral formalism developed here is easily adaptable to handle other types of impurities, such as atomic replacement of individual carbon atoms. Because of the computational efficiency of the code, it is possible to study systems with a large number of dopants in order to investigate the energetics of impurity aggregation and dispersion.

\section{Acknowledgements}
\label{sec:Acknowledgements}
The authors acknowledge the National Research Foundation, Prime Minister's Office, Singapore, under its Medium Sized Centre Programme, and the Ministry of Education, Singapore via grant MOE2017-T2-2-139. A. R. acknowledges support by Yale-NUS College (through grant number R-607-265-380-121). K. N. and S.Y. Q. acknowledge support from Grant NRF-NRFF2013-07 from the National Research Foundation, Singapore. Computations were performed on the NUS CA2DM cluster.

\appendix

\section{Computational Details}
\label{sec:CompDetails}

\begin{figure}
	\includegraphics[width = \columnwidth]{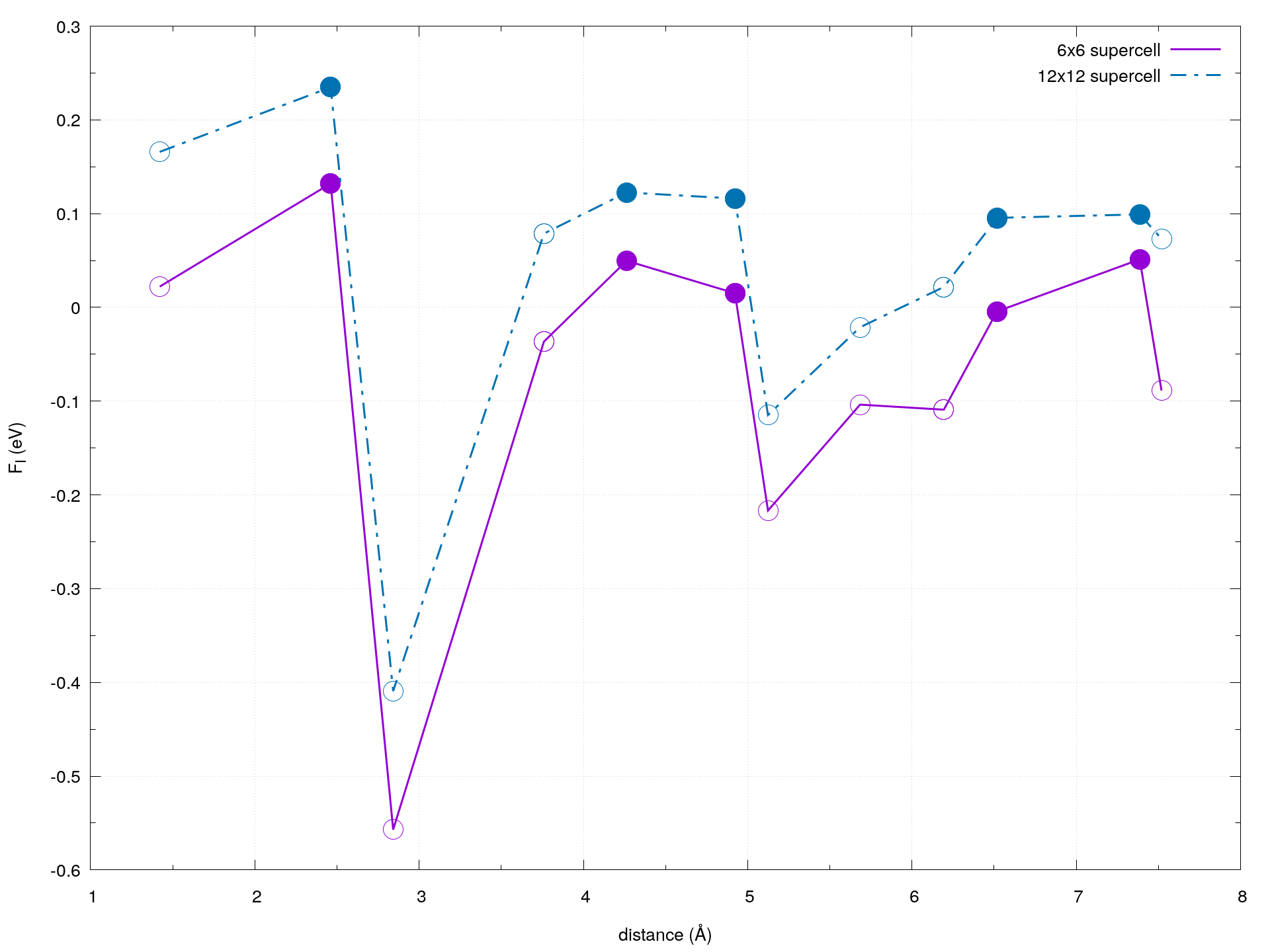}
	\caption{The effect of supercell size on the interaction energy, $F_{I}$, between two H adsorbates. $F_{I}$ is shown for both $6 \times 6$ (solid purple line) and $12 \times 12$ (dashed blue line) graphene supercells, with the H adatoms separated by various distances. Open (closed) circles indicate that the H adatoms reside on the same (opposite) sublattice.}
	\label{fig:66Vs1212Unrel}
\end{figure}

Because of the long-range nature of the graphene-mediated interaction between impurities, the periodic cell considered in our DFT calculations is not, strictly speaking, a system of only two H adatoms; the H atoms in a given cell will feel the interaction of the H atoms in the neighboring cells. The $12 \times 12$ supercell considered in our discussion above, however, is large enough to capture the qualtitative effects of the osicllations in the induced charge and interaction energy. As shown in Fig.~\ref{fig:66Vs1212Unrel}, we see that the shape and magnitude of the interaction energy remains effectively unchanged when the supercell size is reduced to $6 \times 6$.
In a true two-impurity system, as in Sec.~\ref{sec:Analytical_Results}, the interaction energy will go to zero as the impurity separation increases. In the periodic DFT calculations, however, the periodic impurities prevent such a situation and, as a result, an offset is introduced in $F_{I}$. We stress, however, that the fundamental nature of the oscillatory behavior, as well as the scale of the interactions, remains unaffected.

\begin{figure}
	\includegraphics[width = \columnwidth]{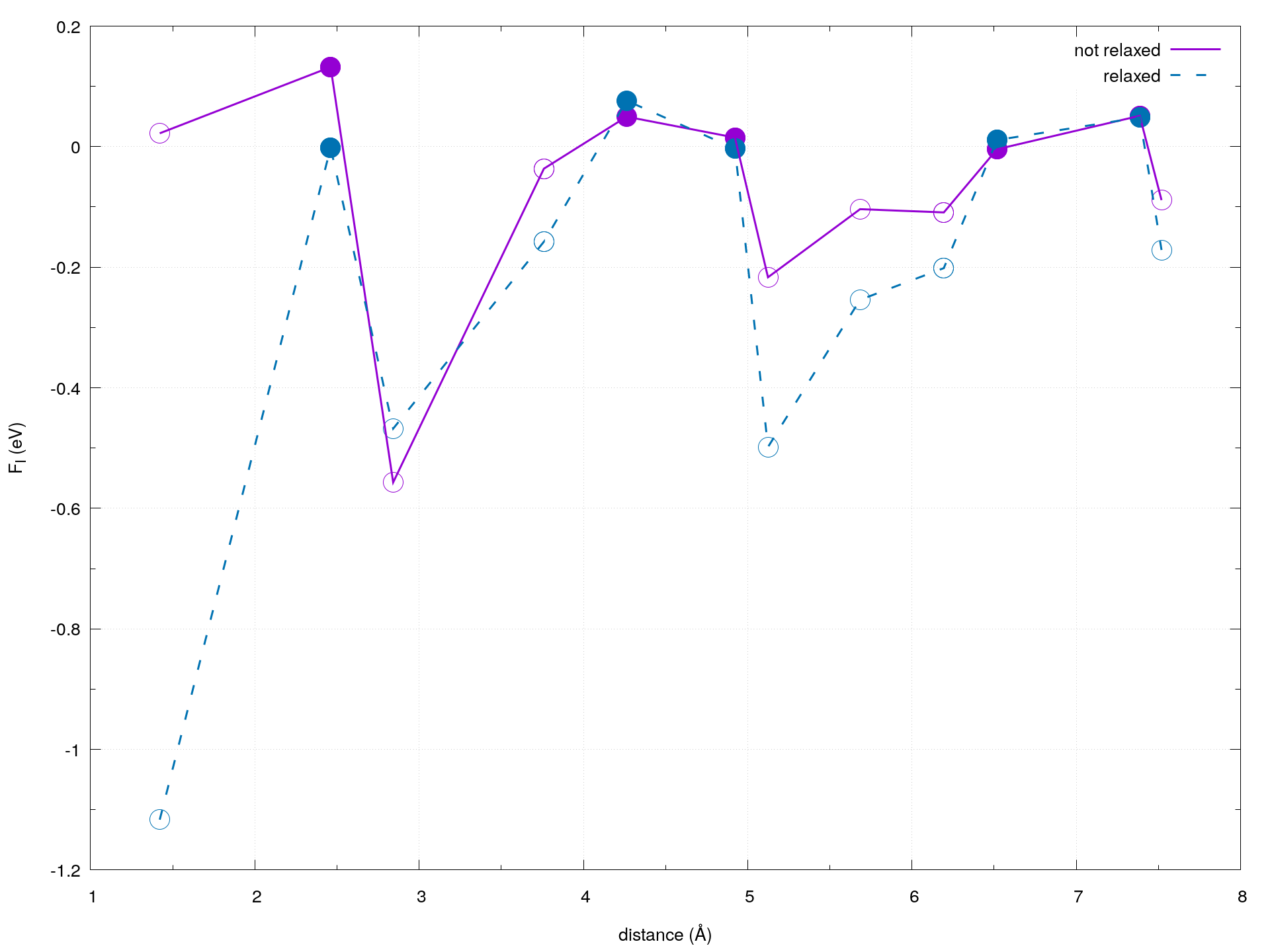}
	\caption{The effect of structural relaxation on the interaction energy, $F_{I}$, between two H adsorbates on a $6 \times 6$ graphene supercell. $F_{I}$ is shown for unrelaxed (solid purple line) and relaxed (dashed green line) systems with the H adatoms separated by various distances. Open (closed) circles indicate that the H adatoms reside on the same (opposite) sublattice. Spin polarization effects are included in all cases.}
	\label{fig:relVsUnrel66}
\end{figure}

We also explore the effects of deformation of the underlying graphene upon adsorption of the H adatoms by structurally relaxing all H-impurity configurations and recomputing $F_{I}$ accordingly. While the graphene lattice undergoes a buckling upon H adsorption, only in the nearest neighbour (smallest separation) configuration is the effect of relaxation meaningful (see Fig.~\ref{fig:relVsUnrel66}).
In all other configurations, magnitude of the interaction energy remains largely unchanged, and the qualitative behavior presented in Fig.~\ref{fig:Interaction:DFT} is preserved.

\begin{figure}
	\includegraphics[width = \columnwidth]{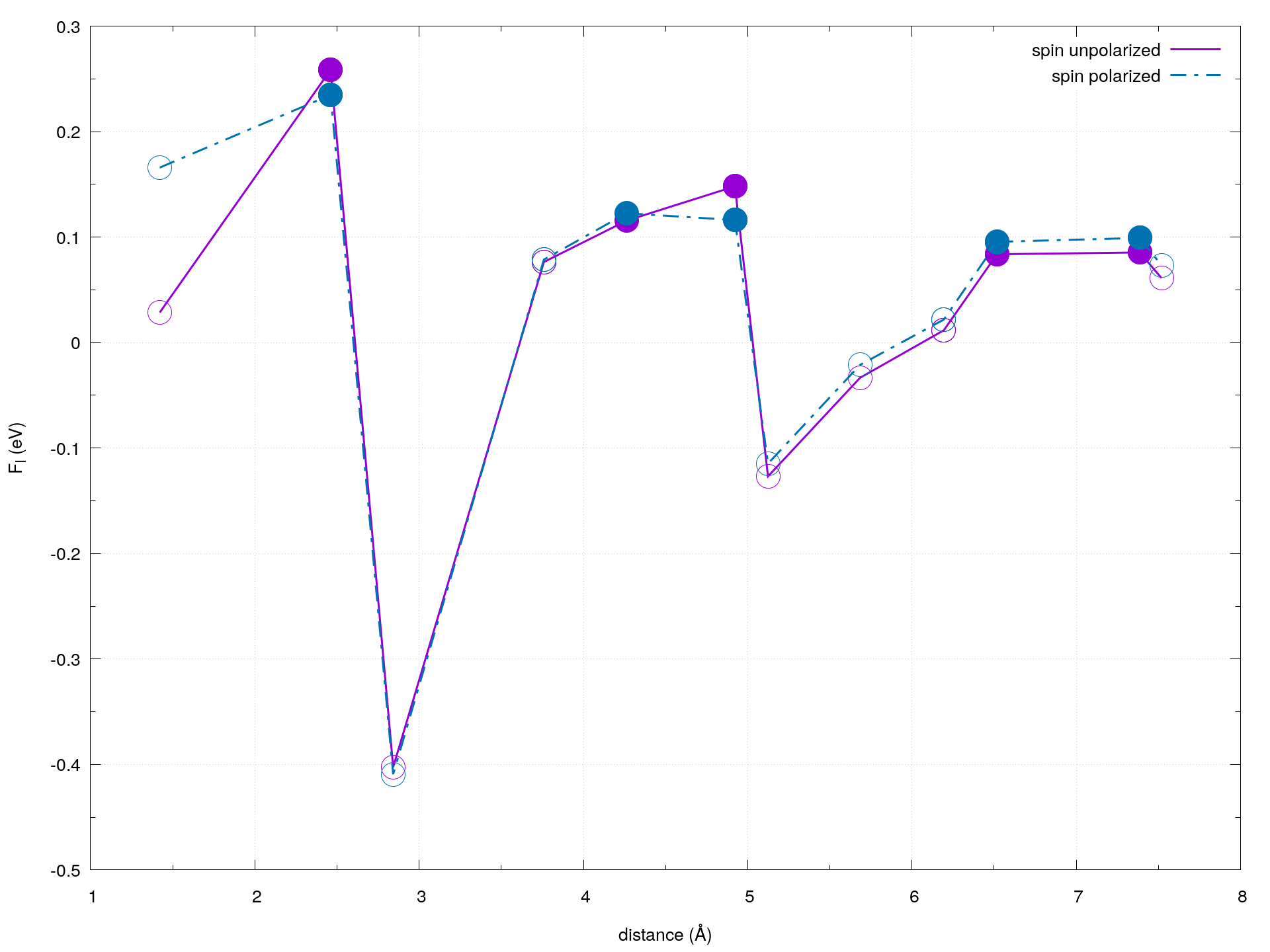}
	\caption{The effect of spin polarization on the interaction energy, $F_{I}$, between two H adsorbates on a $12 \times 12$ graphene supercell. $F_{I}$ is shown for spin unpolarized (solid purple line) and spin polarized (dashed blue line) systems with the H adatoms separated by various distances. Open (closed) circles indicate that the H adatoms reside on the same (opposite) sublattice.}
	\label{fig:spVsNonSp1212}
\end{figure}

Finally, the presence of an H atom introduces an unpaired electron that, in principles, necessiates the inclusion of spin polarization effects. We test the impact of spin polarization by recomputing $F_{I}$ for the undoped $12 \times 12$ supercell (Fig.~\ref{fig:Interaction:DFT}) with spin polarization. As seen in Fig.~\ref{fig:spVsNonSp1212}, the inclusion of spin polarization does not significantly affect $F_{I}$, which justifies its omission in Sec.~\ref{sec:DFT_Results}.

Structural relaxation, total energy and density of states calculations were performed at the DFT level using Quantum ESPRESSO~\citep{Giannozzi2009,Giannozzi2017}. We used the SG15 ONCV~\citep{Hamann2013,Schlipf2015} pseudopotentials with the PBE exchange-correlation functional~\citep{Perdew1996} and a 60 Ry kinetic energy cutoff for wavefunctions. The Brillouin zones of the supercells were sampled using a $36 \times 36 \times 1$ unit cell-equivalent uniform mesh of k-points.

The charge density visualization in Fig.~\ref{fig:Charge:DFT} was produced using VESTA~\citep{Momma2011}.

Numerical integration was performed using the JULIA programming language~\citep{Bezanson2017}. The code can be found at https://github.com/rodin-physics/graphene-QFT.

\section{Graphene Propagator}
\label{sec:Graphene_Propagator}

When dealing with short-range scattering in graphene, the expression
\begin{equation}
	\Omega^{u,v}\left(z\right) =
	\frac{1}{N}\sum_{\mathbf{q}\in\mathrm{BZ}}
	\frac{
		e^{i\mathbf{q}\cdot \left(u\mathbf{d}_1 + v\mathbf{d}_2\right)}
	}
	{z^2 - t^2\left| f_{\mathbf{q}}\right|^2}
	\label{eqn:Omega_R}
\end{equation}
with $ u\mathbf{d}_1 + v\mathbf{d}_2 = \frac{d}{2}\left(u - v, \sqrt{3}\left(u+v\right)\right)$ frequently appears. The momentum integration in Eq.~\eqref{eqn:Omega_R} is complicated by the fact that the Brillouin zone is hexagonal. To make the calculation simpler, we double its area to turn it into a rectangle, which simplifies the integration limits. Using $\mathbf{q}\cdot \left(u\mathbf{d}_1 + v\mathbf{d}_2\right)  = \frac{d}{2}\left[\left(u - v\right)q_x + \sqrt{3}\left(u+v\right)q_y\right]$ and remembering to half the result to account for the increased integration area, we write
\begin{widetext}
	\begin{align}
		\Omega^{u,v}\left(z\right) & = \frac{1}{2}\frac{1}{N}\frac{L^2}{\left(2\pi\right)^2} \int_{-\frac{2\pi}{d}}^\frac{2\pi}{d}dq_x \int_0^\frac{4\pi}{d\sqrt{3}}dq_y
		\frac
		{e^{i \frac{d}{2}\left[\left(u - v\right)q_x + \sqrt{3}\left(u+v\right)q_y\right]}}
		{z^2 - t^2\left|1 + 2\cos\left(\frac{q_xd}{2}\right)e^{i\frac{\sqrt{3}dq_y}{2}}\right|^2}
		\nonumber
		\\
		                           & = \frac{1}{\left(2\pi\right)^2}\oint dx \oint dy
		\frac
		{e^{i \left[\left(u - v\right)x + \left(u+v\right)y\right]}}
		{z^2 - t^2\left(1 + 4\cos^2 x + 4 \cos x\cos y \right)}\,,
		\label{eqn:Omega_R_2}
	\end{align}
\end{widetext}
where $L^2$ is the area of the system. Using
\begin{equation}
	\oint d\theta \frac{e^{il\theta}}{w-\cos\theta} = 2\pi \frac{\left(w - \sqrt{w - 1}\sqrt{w + 1}\right)^{|l|}}{\sqrt{w - 1}\sqrt{w + 1}}\,,
	\label{eqn:Ang_Int}
\end{equation}
we get
\begin{align}
& \Omega^{u,v}\left(z\right) = \frac{1}{2\pi}\frac{1}{4t^2}
	\nonumber
	\\
	\times &
	\oint dx \frac{e^{i\left(u - v\right)x}}{\cos x}\frac{\left(W - \sqrt{W - 1}\sqrt{W + 1}\right)^{|u+v|}}{\sqrt{W - 1}\sqrt{W + 1}}\,,
	\label{eqn:Omega_R_3}
	\\
	       & W = \frac{\frac{z^2}{t^2}-1}{4\cos x}-\cos x\,.
	\label{eqn:W}
\end{align}
From this, $\Xi^{u,v}_{\omega_n}=\Xi^{\mathbf{R}}_{\omega_n}$ for $\mathbf{R} = u\mathbf{d}_1 + v\mathbf{d}_2$ can be written as
\begin{widetext}
	\begin{equation}
		\Xi^{u,v}\left(z\right) =
		\begin{pmatrix}
			z\Omega^{u,v}\left(z\right)
			 &
			- t\left[\Omega^{u,v}\left(z\right) + \Omega^{u+1,v}\left(z\right) + \Omega^{u,v+1}\left(z\right)\right]
			\\
			- t\left[\Omega^{u,v}\left(z\right) + \Omega^{u-1,v}\left(z\right) + \Omega^{u,v-1}\left(z\right)\right]
			 &
			z\Omega^{u,v}\left(z\right)
		\end{pmatrix}\,.
	\end{equation}
\end{widetext}

In the main text, one can see that the function $\Xi^{u_l-u_m,v_l-v_m}\left(z\right)$ appears only as $I_l^\mathrm{T}\Xi^{u_l-u_m,v_l-v_m}\left(z\right) I_m$. This means that it is not necessary to calculate all the elements of $\Xi^{u_l-u_m,v_l-v_m}\left(z\right)$, reducing the computational load.

\bibliography{grapheneImpurityInteraction}

\end{document}